\documentclass[11pt,letterpaper]{article}
\usepackage{doi}
\usepackage[left=1.1in, right=0.9in, top=0.9in, bottom=1in]{geometry}
\usepackage{graphicx}
\usepackage{bm}
\usepackage{amsmath}
\usepackage{amssymb}
\usepackage[dvipsnames]{xcolor}
\usepackage{setspace}
\usepackage[labelfont=bf, font=small]{caption}
\usepackage{abstract}
\usepackage[T1]{fontenc}
\usepackage[round,sort&compress,comma,numbers]{natbib}
\usepackage{hyperref}
\hypersetup{colorlinks=true,
            linkcolor=RubineRed,
            citecolor=RoyalBlue,
            urlcolor=RoyalBlue}

\DeclareCaptionLabelSeparator{colon}{.~}

\begin{document}

\begin{centering}

\textbf{\large The Phase Stability Network of All Inorganic
Materials}\\[0.5\baselineskip]

Vinay~I.~Hegde,$^{\rm 1,}$\footnote{\label{foot:equal}These authors contributed
equally to this work}
Muratahan~Aykol,$^{\rm
2,}$\textsuperscript{\ref{foot:equal},\ref{foot:email}}
Scott~Kirklin,$^{\rm 1,}$\footnote{Current address: Jump Trading LLC, Chicago,
IL 60654} and
Chris~Wolverton$^{\rm 1,}$\footnote{\label{foot:email}Email:
\href{emailto:c-wolverton@northwestern.edu}{c-wolverton@northwestern.edu}
(C.~Wolverton), \href{emailto:murat.aykol@tri.global}{murat.aykol@global.tri}
(M.~Aykol)}\\[0.5\baselineskip]

$^{\rm 1}$\textit{Department of Materials Science and Engineering,\\
Northwestern University, Evanston, IL 60208}\\
$^{\rm 2}$\textit{Toyota Research Institute, Los Altos, CA
94022}\\[\baselineskip]

\end{centering}

\clearpage

\begin{abstract}
\textbf{One of the holy grails of materials science, unlocking
structure-property relationships, has largely been pursued via bottom-up
investigations of how the arrangement of atoms and interatomic bonding in a
material determine its macroscopic behavior. Here we consider a complementary
approach, a top-down study of the organizational structure of networks of
materials, based on the interaction between materials themselves. We unravel the
complete ``phase stability network of all inorganic materials'' as a
densely-connected complex network of 21,000 thermodynamically stable compounds
(nodes) interlinked by 41 million tie-lines (edges) defining their two-phase
equilibria, as computed by high-throughput density functional theory. We find
that the node connectivity in the materials network has a lognormal
distribution, and the connectivity decreases with the number of elemental
constituents in a material. Analyzing the topology of this network of materials
has the potential to uncover new knowledge inaccessible from traditional
atoms-to-materials paradigms. Using the connectivity of nodes in the phase
stability network, we derive a rational, data-driven metric for material
reactivity, the ``nobility index'', and quantitatively identify the noblest
materials in nature.}
\end{abstract}

\section*{Introduction}\label{sec:intro}

Several diverse complex systems are modeled as networks of discrete
components linked together: man-made systems such as electrical power grids and
the world-wide web~\cite{watts_collective_1998, albert_diameter_1999},
social systems such as friendship and scientific
collaborations~\cite{watts_identity_2002, newman_structure_2001}, and
natural systems such as metabolism in a cell and
food-webs~\cite{williams_simple_2000, guimera_functional_2005}. Despite
significant variation in the nature of individual components and
interconnections, many of these networks show striking similarities in their
topology~\cite{barabasi_emergence_1999, strogatz_exploring_2001}, often
providing new insights into each respective domain of knowledge. For instance,
disparate systems such as the world-wide web and metabolic reactions in cellular
organisms both have been shown to follow the organizational principles of
robust, error-tolerant scale-free networks, with implications for the resilience
of the internet and the design of therapeutics~\cite{jeong_large-scale_2000,
strogatz_exploring_2001}, respectively.

Recent developments in high-throughput density functional theory
(HT-DFT)~\cite{curtarolo_high-throughput_2013} have resulted in massive
computational databases of materials properties~\cite{saal_materials_2013,
kirklin_oqmd_2015, Jain2013, curtarolo_aflowlib.org_2012, Hachmann2011},
containing the calculated properties of hundreds of thousands of experimentally
reported and hypothetical materials. Such databases have led to new data-driven
approaches toward understanding materials. Here we introduce a novel paradigm of
viewing materials, and equilibrium phase diagrams in particular, via the lens of
complex network theory, i.e.\ studying the similarities and interactions between
materials themselves, in striking contrast to the traditional bottom-up
approaches toward unlocking structure-property relationships in
materials~\cite{dove2003structure, phillips2001crystals}.

\begin{figure}[!ht]
\centering
\includegraphics[width=0.95\textwidth]{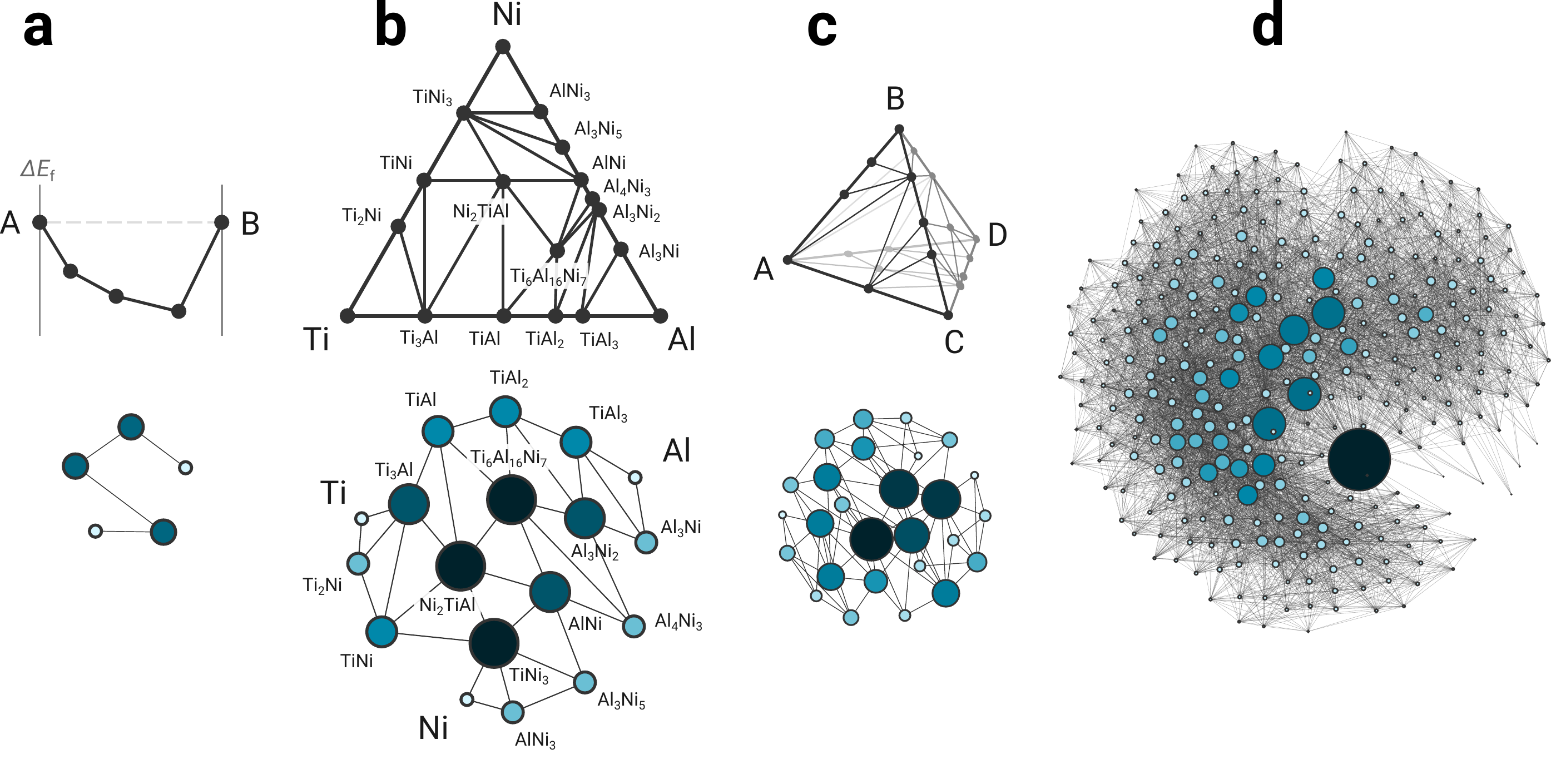}
\caption{\textbf{Network representation of \textit{T}=0~K materials phase
diagrams.} Stable phases and two-phase equilibria (tie-lines) in a phase diagram
are represented as nodes and edges, respectively, to create the corresponding
network: (\textbf{a}) Schematic A-B binary system represented as a typical
two-dimensional convex hull of compound formation energies. (\textbf{b})
Ti-Ni-Al as an example ternary system, with the $T$=0~K phase diagram shown as a
Gibbs triangle. (\textbf{c}) Schematic A-B-C-D quaternary phase diagram shown as
a Gibbs tetrahedron. (\textbf{d}) The 3$d$ transition metal-chalcogen (i.e.\
14-dimensional chemical space) materials network. No conventional visual
representations exist of phase diagrams at higher than 4 dimensions. Node sizes
shown are proportional to node degree.}\label{fig:schematic}
\end{figure}

We use the Open Quantum Materials Database (OQMD)~\cite{saal_materials_2013,
kirklin_oqmd_2015}, a HT-DFT database containing calculations of nearly all
crystallographically ordered, structurally unique materials experimentally
observed to date (as collected in the Inorganic Crystal Structure
Database~\cite{belsky_new_2002} repository) and a large number of hypothetical
materials constructed using commonly occurring structural prototypes---a total
of more than half a million materials---to extract the ``universal phase stability
network'' or the ``universal $T$=0~K phase diagram''. We accomplish this by
using all the phase data in the OQMD within a convex-hull formalism, and
identifying \textit{all} thermodynamically stable materials and \textit{all}
two-phase equilibria between them. We then represent stable materials as nodes
and two-phase equilibria (tie-lines) as edges, thus describing a $T$=0~K phase
diagram as a network encoding thermodynamic stability (illustrated with
schematics in Fig.~\ref{fig:schematic}).

\section*{Results}\label{sec:results}

\subsection*{Overall network connectivity}\label{ssec:network_connectivity}

We find that the phase stability network of all inorganic materials consists of
$\sim$21,300 nodes and is remarkably dense with a total of nearly 41 million
edges, and extremely well-connected with $\sim$3,850 edges per node on average
(``mean degree'' $\langle k \rangle$). This means that every stable inorganic
compound can form a stable two-phase equilibrium with 3,850 other compounds on
average. For comparison, $\langle k \rangle$ for other widely-studied networks
range from 1.4 (network of email messages) to 113.4 (collaboration network of
film actors)~\cite{newman_structure_2003}. The connectance of the materials
network, or the fraction of the maximum possible number of edges that are
actually present is 0.18. This is an important statistic for the design of
``systems of materials'', such as electrodes and electrolytes making up
batteries~\cite{ong_phase_stability_2013}, or coating materials separating
two reactive components~\cite{aykol_ht_coatings_2016}, where the longevity
of the system relies on stable coexistence of such components. Using a
lithium-ion intercalation battery as an example ``system of materials'', a
common approach to tackling electrode degradation is to apply protective
coatings on electrode particles. In such a battery, the material in the
electrode coating should not react with/be consumed by materials in the
electrode \textit{as well as} those in the
electrolyte~\citep{aykol:ncomm2016, snydacker:jecs2017}. Thus, the
coating--electrode and the coating--electrolyte material pairs must both have
tie-lines with each other in order to stably coexist in the system. In other
words, both pairs must be neighboring, connected nodes in the materials network.

The degree distribution in the complete phase stability network, specifically
the probability $p(k)$ that a material has a tie-line with $k$ other materials
in the network follows a lognormal form (Fig.~\ref{fig:topology}a, and Fig.~1
in the Supplementary Materials). While many widely-studied networks are known to
have scale-free power-law degree distributions, lognormal distributions are
another member of the ``heavy-tail'' family, are also relatively common, and
behave quite similar to power-laws~\cite{mitzenmacher_brief_2003}. In fact,
sparsity has been shown to be a necessary condition for the emergence of an
exact power-law behavior~\cite{delgenio_all_2011}, and densification in sparse,
scale-free networks leads to distributions that deviate from a power-law and
become closer to lognormal. Thus, the lognormal behavior of the materials
network can be understood to result from its extremely dense connectivity, in
contrast to the general sparsity of commonly-studied networks.

\begin{figure}[!htbp]
\centering
\includegraphics[width=0.90\textwidth]{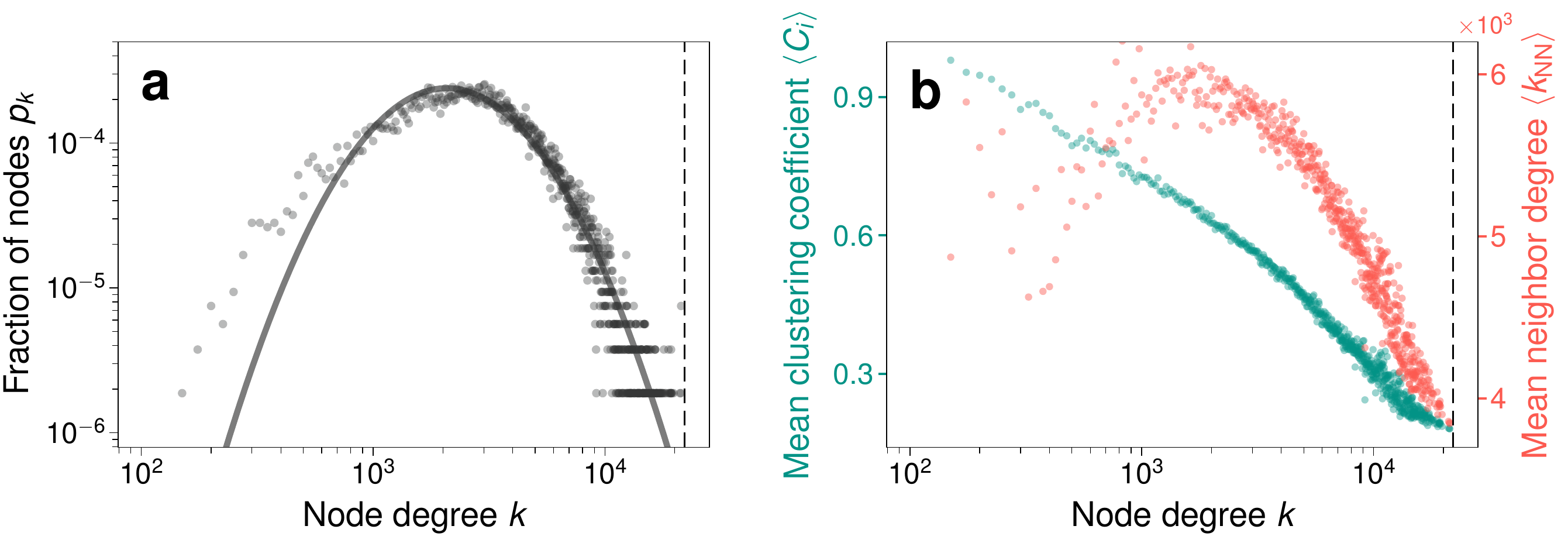}
\caption{\textbf{Overall structure and topology of the materials network.}
(\textbf{a}) The distribution of node degree in the materials network (grey
circles) shows a heavy tail, i.e.\ a sizeable fraction of materials have
tie-lines with nearly all other materials. A lognormal fit is shown as a solid
grey line. (\textbf{b}) The mean local clustering coefficient $\langle
\mathcal{C}_i \rangle$ (green) decreases with node degree $k$ indicating that
stable materials form local, high-connected communities. The mean neighbor
degree $\langle k_{\rm NN} \rangle$ (red) also decreases with $k$, implying a
weakly-dissortative network behavior, i.e.\ materials with a large number of
tie-lines connect with those with fewer tie-lines in the network. In both
subplots, the vertical dashed line represents the total number of nodes (stable
materials) in the network.}\label{fig:topology}
\end{figure}

\subsection*{Network topology}\label{ssec:network_topology}

The characteristic path length or mean node-node distance in a network,
$\mathcal{L}$, is defined as the number of edges in the shortest path between
two nodes, averaged over all pairs of nodes. The longest node-node distance in
the network defines its diameter, $\mathcal{L}_{\rm max}$. The characteristic path
length of the materials network $\mathcal{L} = 1.8$, and its diameter
$\mathcal{L}_{\rm max} = 2$. This remarkably short path length indicates that the
materials network has ``small-world''
characteristics~\cite{watts_collective_1998}, i.e.\ despite its large
size, the number of edges that need to be traversed from a given node to any
other node is relatively small. The extremely small $\mathcal{L}$ for the
materials network can be intuitively understood to be a consequence of the
almost complete lack of reactivity of noble gases. The non-participation of
noble gases in the formation of compounds (and thus having tie-lines with nearly
all materials in the network) places an upper bound of 2 on $\mathcal{L}_{\rm max}$,
and since some material pairs already have tie-lines that connect them
immediately, the mean path $\mathcal{L}$ is slightly smaller than 2. Even if
noble gases are disregarded, the mean path length and diameter of the materials
network remain small due to the presence of a few other very-highly connected
nodes corresponding to extremely stable and non-reactive materials, e.g.\ binary
halides.

Another metric of interest in a real-world network is transitivity or
clustering, quantified by its clustering coefficient, $\mathcal{C}$, which is
the probability that two nodes connected to the same third node are themselves
connected. In other words, given that there exist stable two-phase equilibria
A--C and B--C, what is the probability that A and B can stably coexist?
Depending on how the averaging is performed, a global ($\mathcal{C}_g$) or mean
local ($\bar{\mathcal{C}_i}$) cluster coefficient of a network can be
defined~\cite{newman_structure_2003, watts_collective_1998}.
For the materials network, the clustering coefficients are $\mathcal{C}_g =
0.41$ and $\bar{\mathcal{C}_i} = 0.55$, comparable to other real-world networks,
and much higher than random networks of the same density. The mean local
clustering coefficient of the materials network decreases with increasing node
connectivity (Fig.~\ref{fig:topology}b), indicating that stable materials form local
highly-connected communities in the network, and such behavior often suggests a
hierarchical network structure~\cite{ravasz_hierarchical_2003}. The
assortativity coefficient or the Pearson correlation coefficient of degree
between pairs of connected nodes in the materials network is $-0.13$, indicating
weakly dissortative mixing behavior. This is also confirmed by the distribution
of the mean degree of neighbors of a node of degree $k$ being a decreasing
function of $k$ (Fig.~\ref{fig:topology}a). In other words, materials with a
high $k$ (i.e.\ large number of tie-lines) tend to connect with materials with a
lower $k$ (i.e.\ smaller number of tie-lines). This weakly dissortative behavior
of the materials network is similar to that observed in most other
technological, information, biological networks, and is likely a virtue of such
networks being simple graphs~\cite{maslov_detection_2004}.

\subsection*{Hierarchy in the materials network}\label{ssec:hierarchy}

\begin{figure}[!htbp]
\centering
\includegraphics[width=0.90\textwidth]{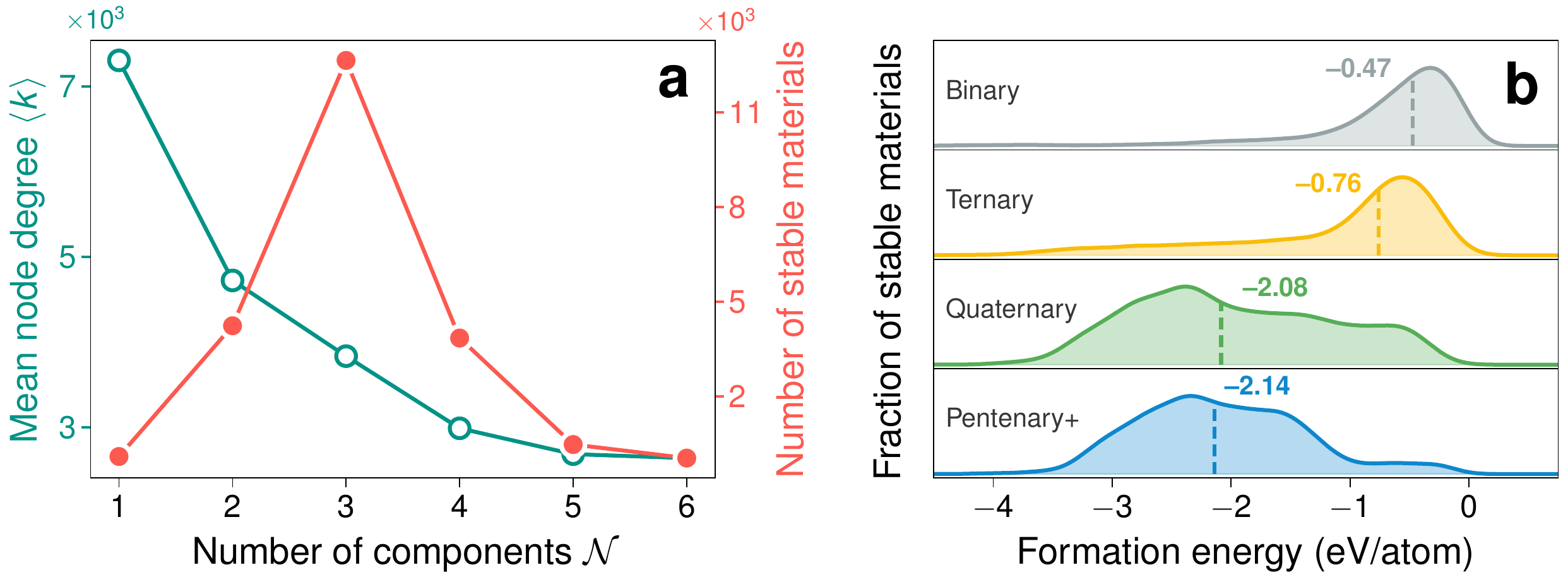}
\caption{\textbf{Hierarchy in the materials network, and underlying energetic
considerations.} (\textbf{a}) The mean node degree or average number of tie-lines
$\langle k \rangle$ (green, open) decreases as a function of number of
components $\mathcal{N}$ (i.e.\ binary, ternary, and so on), which results from
high-$\mathcal{N}$ materials having to compete with low-$\mathcal{N}$ materials
for stability. The number of known stable $\mathcal{N}$-ary materials (red)
itself actually peaks at $\mathcal{N}=3$ (ternaries). (\textbf{b}) Gaussian
kernel density estimates of compound formation energies for all stable materials
separated by number of components in the material. Dotted vertical lines
indicate the respective median of each distribution. High-$\mathcal{N}$ need
significantly lower formation energies than low-$\mathcal{N}$ materials to
become stable, e.g.  $-2.08$ versus $-0.47$~eV/atom for quaternary and binary
materials, respectively.}\label{fig:hierarchy}
\end{figure}

The mean degree or the average number of tie-lines per material $\langle k
\rangle$ decreases with the number of components, $\mathcal{N}$ ($\mathcal{N} =
2$ for binary, $\mathcal{N} = 3$ for ternary, etc. See
Fig.~\ref{fig:hierarchy}a), indicating a
chemical hierarchy in the materials network. This can be understood to result
from an inherent competition for tie-lines that high-$\mathcal{N}$ materials
face with low-$\mathcal{N}$ materials in their chemical space, but not
vice-versa. In other words, ternary compounds $X_aY_bZ_c$ compete not only with
other compounds in the $X$-$Y$-$Z$ chemical space but also with binary compounds
in the $X$-$Y$, $Y$-$Z$, $Z$-$X$ spaces for tie-lines.

We note that this decrease in $\langle k \rangle$ with $\mathcal{N}$ is distinct
from the distribution of number of stable $\mathcal{N}$-ary materials itself
(Fig.~\ref{fig:hierarchy}a), which shows a peak at $\mathcal{N} = 3$. Does
this peak in the distribution of stable materials imply the existence of
infinite, underexplored space for the discovery of new materials beyond
ternaries? The distribution of formation energies of materials as a function of
number of components $\mathcal{N}$ (Fig.~\ref{fig:hierarchy}b) reflects the
consequence of competition between low- and high-component materials:
high-$\mathcal{N}$ compounds appear to need significantly lower formation
energies than low-$\mathcal{N}$ ones to become stable. Since there is no obvious
underlying reason for the distribution of $T=0$ K formation energies (with
entropic effects neglected) to differ significantly with $\mathcal{N}$, only a
few high-$\mathcal{N}$ materials can ``survive'' as stable phases if the
corresponding lower-$\mathcal{N}$ systems already have several stable phases.
This is consistent with the recent reports of a ``volcano plot'' that emerges
for stable inorganic ternary nitrides as a function of energetic competition
with their corresponding binary nitrides~\cite{sun2019nature}, and an increased
probability of phase separation with increasing number of components in a
material system~\cite{sun2016thermodynamic}. Widom~\cite{widom_frequency_2017}
further argued that the a peak near $\mathcal{N} = 3$ or $4$ in such
distributions arises from a competition between combinatorial explosion and
diminishing volume-to-surface ratio in the composition simplex, as
$\mathcal{N}$ increases.  Thus, although we do not know of a fundamental law
limiting access to thermodynamically stable materials with higher components, a
combination of the hierarchy observed in the phase stability network, the
distribution of formation energies, and the topology of the convex energy
surface all suggest that the scarcity of known high-$\mathcal{N}$ stable
materials is not merely a consequence of those chemical spaces being
underexplored.

\subsection*{Knowledge extraction: material nobility index}\label{sec:nobility}

Since the phase stability network practically encompasses all known inorganic
crystalline materials as well as a large number of predicted hypothetical
materials, the number of tie-lines emerges as a natural metric of nobility of a
crystalline material---it is simply the count of other materials it is determined
to have no reactivity against. Thus, while material reactivity or nobility have
no standard definitions, a network representation of materials enables us to
tackle the chemical nobility of inorganic materials in solid-solid and solid-gas
reactions in a completely data-driven fashion, instead of the traditional
intuitive or heuristic approaches. Since the number of tie-lines in the
materials network is lognormally distributed, we devise a new standard score of
material nobility, the ``nobility index'':

\begin{equation}\label{eqn:nobility_index}
\mathcal{Z}_n = \frac{\ln(k) - \mu}{\sigma}
\end{equation}

where $k$ is the node degree or the number of tie-lines a material has, and
$\mu = 8.06$ and $\sigma = 0.65$ are the mean and standard deviation of the
underlying lognormal distribution. The nobility index is thus agnostic of
textbook classifications such as metal, nonmetal, metalloid, ionic, covalent,
and so on, and works equally well for any given material. Since the tie-lines in
the network are as computed with DFT, the nobilities of materials predicted
herewith are only limited by DFT accuracy in estimating relative stabilities of
inorganic materials~\cite{kirklin_oqmd_2015, sun2016thermodynamic,
aykol_thermodynamic_2018}.

\begin{figure}[!ht]
\centering
\includegraphics[width=0.90\textwidth]{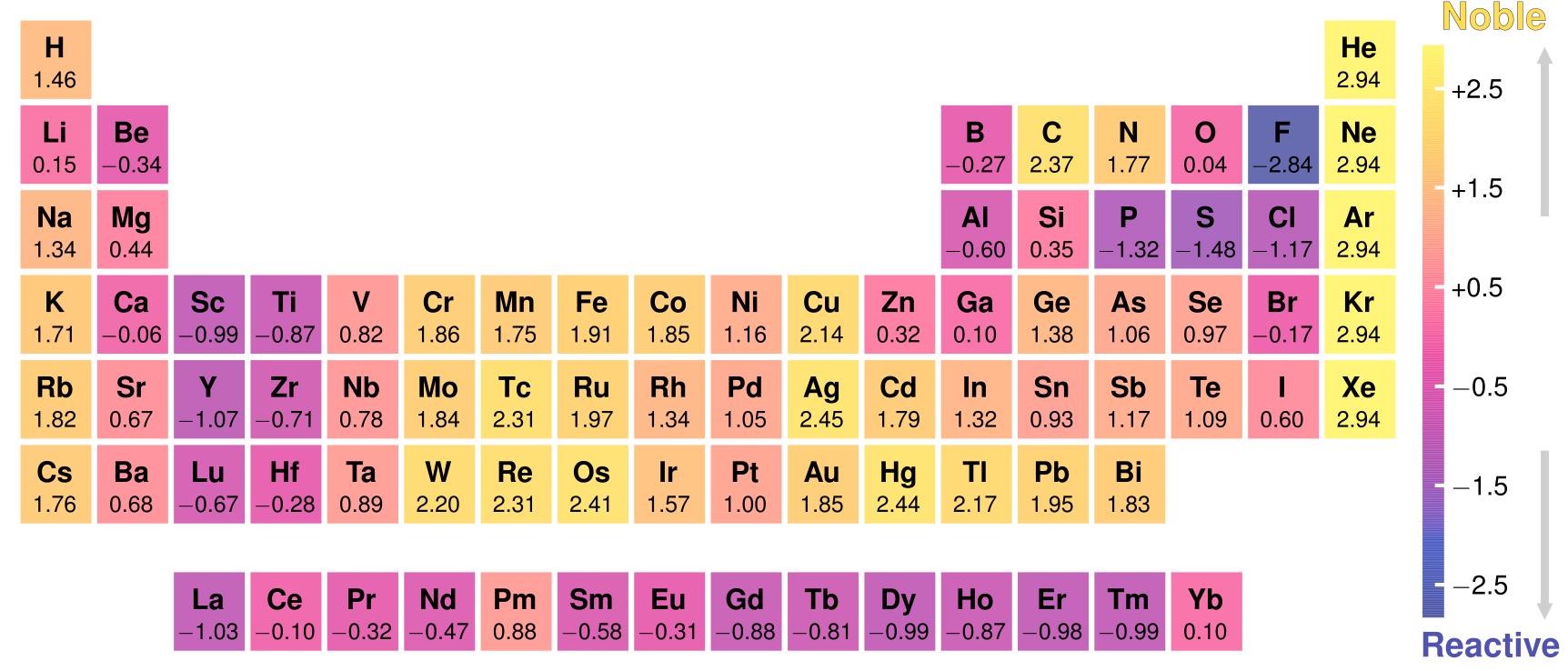}
\caption{\textbf{Nobility index of all elements.} The standard score,
$\mathcal{Z}_n$, derived in this work using material connectivity in the phase
stability network, as a measure of nobility against solid-solid and solid-gas
reactions. Nobility increases up the scale. Numerical values of elemental
$\mathcal{Z}_n$ are given below the respective
symbols.}\label{fig:nobility_index}
\end{figure}

First, we tackle the reactivity or nobility of elements. Noble gases and
fluorine form the bounds of the nobility index
(Fig.~\ref{fig:nobility_index}), as the noblest and the most reactive,
respectively, not only among the elements but in fact among all materials in the
network. The most reactive elements following F are P, S, and Cl. Alkali and
alkaline earth metals, often considered to be highly reactive metals, are
relatively noble in solid-solid and solid-gas reactions, in comparison to early
$d$-block or lanthanide elements, which are, along with Al, the most reactive
metals. The nobility index increases down a group for metals, and increases
(decreases) from left-to-right along a row of the periodic table within the
$d$-block ($s$-block). But what is the noblest metal of them all? Ag emerges as
the noblest of all elements after noble gases, followed closely by Hg, Os, Re,
W, and Cu, all having more than 14,000 tie-lines. Gold, traditionally considered
the noblest element~\cite{hammer_why_1995}, despite being relatively densely
connected with 10,000 tie-lines, is less noble in solid-state reactions.
Finally, we find that $\mathcal{Z}_n$ is not correlated with other common
elemental properties such as electronegativity, atomic radii, melting point, and
others~\cite{Ward2015}, indicating that the nobility index encodes new
information not readily captured by those properties (Fig.~2 in the
Supplementary Materials).

Beyond elements, what are the noblest inorganic compounds of all? The compounds
at the top of the nobility list are IA/IIA-VIIA compounds such as LiF, NaCl, KCl,
CsCl, KBr, CsBr, KI, RbI, CaF$_2$, SrF$_2$, CsYbF$_3$, RbYbF$_3$, and others,
their inertness likely due to stability from strong ionic bonding between their
constituents. We exclude rareearth- and actinide-containing compounds from the
previous analysis of compound nobility in order to account for any shortcomings
in the DFT description of $f$-block elements and compounds containing them.

\section*{Discussion}\label{sec:discussion}

While some of our findings above are in line with chemical intuition, relative
nobilities in certain cases, e.g.\ silver vs gold, deviate from it. This
deviation is in part due to the historical context in which these materials have
been considered noble or reactive, e.g.\ whether an element oxidizes or corrodes
readily in air, reacts with water and/or certain acids, dissolves in water or
electrolytes, and how vigorous such reactions seem. More fundamental approaches
to finding descriptors for reactivity go back to electronegativity related
concepts, followed by interrelated theories based on perturbation theory,
derivatives of electronic energy such as hardness and softness, and
others largely developed for molecules~\cite{klopman_chemical_1968,
yang_hardness_1985, chermette_chemical_1999}. In contrast, the nobility index,
$\mathcal{Z}_n$, as derived from the tie-lines in the network of all inorganic
materials represents a general metric emerging directly from bulk thermodynamic
data.

High-throughput experimental and computational techniques are leading to an
explosive growth in the size of materials databases. Representation and
interpretation of the data at a large-scale, however, remains a challenge. Here
we show that tools from complex network theory enable us to access
otherwise-difficult-to-extract information from such large datasets. In other
words, the emergence of material reactivity from the collective behavior of all
materials in the phase stability network serves as a simple, preliminary example
of knowledge extraction out of complex networks of materials. Other similar
approaches can be used to discover other hidden knowledge, e.g. analysis of
``communities'' or ``cliques'' in the network of all materials can uncover
hitherto-unknown relationships between various known materials.

Further, there are various ways our graph theoretic approach to materials data
can be used to immediately applied to new materials discovery and design: (a)
\textit{direct} techniques, e.g.\ metrics from network theory such as local
clustering and similarity can be used to identify ``holes'' in the current
network---where nodes (i.e.\ materials) are expected to exist but currently
do not, and (b) \textit{indirect} techniques, e.g.\ using the extracted
knowledge or quantities derived from the network as input to other approaches
such as in materials informatics. For example, using temporal materials
discovery information in combination with thermodynamic phase stability networks
can help predict synthesizability~\cite{aykol_network_2019}.  Furthermore, while
some of its features resemble other complex networks, the extremely-high
connectance and the lognormal degree distribution of the presented phase
stability network imply that its underlying generative mechanisms may be unique,
and developing generative models for such materials networks can have
significant impact on the knowledge discovery of materials in the future.

\section*{Methods}\label{sec:matmethods}
All convex hull constructions were performed using the \textsf{\textit{Qhull}}
library~\cite{Barber96thequickhull} as implemented in the \textsf{\textit{qmpy}}
(\href{https://pypi.org/project/qmpy}{pypi.org/project/qmpy}) package. All
network analyses were performed using the
\textsf{\textit{graph-tool}}~\cite{peixoto_graph-tool_2014} and
\textsf{\textit{powerlaw}}~\cite{alstott_powerlaw:2014} packages, and
comparison of heavy-tailed distributions is done according to the method of log
likelihood ratios as described in Clauset~et~al.~\cite{clauset2009}. Details of
the divide-and-conquer approach used to tackle the combinatorial explosion in
calculating the universal phase diagram, the related exponential increase
in the time complexity to construct convex hulls in higher
dimensions~\cite{Sartipizadeh2016}, its network representation, and determining
the node degree distribution are provided in the Supplementary Materials.\\

\begin{spacing}{1.0}
\bibliographystyle{ScienceAdvances}
\bibliography{references}
\end{spacing}

\section*{Acknowledgements}\label{sec:ack}
\textbf{Funding:} V.I.H. acknowledges support from Toyota Research Institute
(TRI) through the Accelerated Materials Design and Discovery program. C.W.
acknowledges the support of the National Science Foundation (NSF), through the
MRSEC program, grant number DMR-1720139. \textbf{Author contributions:} V.I.H.
and M.A. conceived and designed the project, and contributed equally to this
work. M.A. calculated all the tie-lines in the materials network. V.I.H.
performed the network analysis and nobility index calculations. S.K. wrote the
code to calculate convex hulls. C.W. supervised the project. All authors
contributed toward writing the manuscript. \textbf{Competing interests:} None.
\textbf{Data availability:} All data needed to evaluate the conclusions in the
paper are present in the paper, the Supplementary Materials, or is available
to download at no cost from the OQMD website
(\href{http://oqmd.org}{http://oqmd.org}). Additional data related to this
paper may be requested from the authors.

\section*{Supplementary Materials}\label{sec:suppmat}
Supplementary Text (divide-and-conquer calculation of the universal
phase diagram, fitting degree distributions, and novel materials knowledge
captured by the nobility index)\\
Figs.~S1--S3\\
Table~S1 with compute times for determining tie-lines\\
References

\end{document}


\begin{centering}

\textbf{\Large Supplementary Materials}\\[1.0\baselineskip]

\textbf{\large The Phase Stability Network of All Inorganic
Materials}\\[0.5\baselineskip]

Vinay~I.~Hegde,$^{\rm 1,}$\footnote{\label{foot:equal}These authors contributed
equally to this work}
Muratahan~Aykol,$^{\rm
2,}$\textsuperscript{\ref{foot:equal},\ref{foot:email}}
Scott~Kirklin,$^{\rm 1,}$\footnote{Current address: Jump Trading LLC, Chicago,
IL 60654} and
Chris~Wolverton$^{\rm 1,}$\footnote{\label{foot:email}Email:
\href{emailto:c-wolverton@northwestern.edu}{c-wolverton@northwestern.edu}
(C.~Wolverton), \href{emailto:murat.aykol@tri.global}{murat.aykol@global.tri}
(M.~Aykol)}\\[0.5\baselineskip]

$^{\rm 1}$\textit{Department of Materials Science and Engineering,\\
Northwestern University, Evanston, IL 60208}\\
$^{\rm 2}$\textit{Toyota Research Institute, Los Altos, CA
94022}\\[\baselineskip]

\end{centering}

\clearpage

We describe
the approach used to calculate the universal phase diagram, its network
representation and determining the node degree distribution in the network in
the following sections S1--S2.\\

\noindent
\textbf{\large S1. Calculation of the $\mathbf{T=0}$~K universal phase
diagram}\\[-0.75\baselineskip]

The $T=0$~K phase diagram for a given chemical space is determined by the
so-called convex hull construction. A phase is thermodynamically stable iff it
lies on (i.e.\ is a vertex of) the convex hull of $T=0$~K formation energies of
all phases in the chemical space. And phases that are directly connected by a
tie-line, i.e., phases that lie on the same facet of the convex hull, are in
equilibrium with one another.  Determining a binary A-B phase diagram requires
constructing a 2-dimensional convex hull of formation energies of all
A$_{x}$B$_{y}$ compounds (composition $x$ and formation energy being the two
dimensions), a ternary A-B-C phase diagram requires constructing a 3-dimensional
convex hull of formation energies of all A$_{x}$B$_{y}$C$_{z}$ compounds
(compositions $x$ and $y$, and formation energy being the three dimensions), and
so on. The determination of an $d$-nary phase diagram requires the construction
of an $d$-dimensional convex hull of formation energies of all the $N$ phases in
the chemical space.

For low dimensions, i.e.\ $d = 2$ or 3 (binary or ternary systems), finding the
convex hull of $N$ points (total number of phases) has a worst-case time
complexity of $\mathcal{O}(N\,log\,N)$. For higher dimensions, standard methods
of determining convex hulls such as the Quickhull
algorithm, have worst-case time complexities of
$\mathcal{O}(N^{[d/2]})$~\cite{Barber96thequickhull}. For random data, even the
average-case time complexity at higher dimensions scales as
$\mathcal{O}(log^{d-1}N)$, i.e.\ exponentially with $d$~\cite{Sartipizadeh2016}.
Such scaling behaviors mean that for moderately large number of points $N$ and
dimensions $d$, finding the convex hull becomes increasingly practically
challenging. For instance, to find the convex hull of all known inorganic
materials, even restricting ourselves to experimentally reported compounds in
the Open Quantum Materials Database (OQMD), $N \approx 40,000$ and $d = 89$,
making the calculation of the convex hull practically impossible with a
traditional single-shot approach.

We tackle this challenge of calculating the convex hull at high-dimensions by
using a divide-and-conquer approach. While the representational complexity of
the convex-hull increases at least exponentially with $d$, we know from the set
of existing materials that not many of them are high-dimensional by themselves.
In fact, 99.5\% of materials in the OQMD have 4 unique elemental components or
fewer. Since the stability of a material is determined only within the chemical
subspace of elements that it is made of, we first determine the vertices (i.e.\
stable materials) of the 89-dimensional convex hull at a reduced computational
cost by computing the convex hulls in low-dimensional subspaces for each
individual material separately. For instance, to determine if the compound
CaMnO$_3$ is stable, it is sufficient to construct the convex hull of all phases
Ca$_x$Mn$_y$O$_z$ in the Ca-Mn-O subspace. This process of constructing convex
hulls separately for each unique chemical subspace yields all the vertices of
the full convex hull: $\sim\! 2.1 \times 10^{4}$ stable materials out of the
$>\! 5.5 \times 10^{5}$ total materials calculated in the OQMD\@. Having
determined the vertices of the full convex hull, in the second stage, we
exhaustively evaluate the existence of a tie-line between any given pair of
stable compounds in the OQMD by constructing the convex hull of formation
energies in the combined chemical spaces of such candidate pairs, rather than
the full 89-dimensional space itself. For example, to determine whether there
exists a tie-line between Li$_2$O and NaCl, we construct the Li-Na-Cl-O convex
hull, and find that there indeed exists a Li$_2$O-NaCl tie-line. In contrast,
from a Na-K-F-Cl convex hull we find that NaCl and KF, in fact, ``react'' to
form a NaF-KCl two-phase equilibrium. Overall, we construct convex hulls for all
$^{2.1 \times 10^4}C_2 \approx 2.3 \times 10^8$ stable phase combinations, and
find a total $\sim\!41 \times 10^6$ tie-lines.

The computational cost of constructing a convex hull for a unique chemical
subspace is expectedly highly dependent on the number of components, and ranges
from a few seconds to a few minutes on a standard desktop computer utilizing a
single core (some sample times for checking if a tie-line exists between two
known materials are provide in Table~\ref{tab:sample_times_ch}). With a
conservative estimate of 15--20 seconds per tie-line, the total time required to
exhaustively evaluate all possible tie-lines is more than 1 million CPU hours.
\begin{table}
\centering
\begin{tabular}{llcc}
    \toprule
    Phase 1 & Phase 2 & \# Components & Time (s) \\
    \midrule
    Na$_2$O & KCl & 4 & $\sim$3 \\
    Fe$_2$S$_3$ & Li$_2$MnO$_4$ & 5 & $\sim$6 \\
    Li$_3$PS$_4$ & SrTiO$_3$ & 6 & $\sim$8 \\
    Ba$_2$Li$_3$TaN$_4$ & LiCoO$_2$ & 6 & $\sim$14 \\
    Ba$_2$Li$_3$TaN$_4$ & NaCoO$_2$ & 7 & $\sim$32 \\
    Mn$_2$Hg$_2$SF$_6$ & Li$_4$CrCoO$_6$ & 8 & $\sim$34 \\
    Mn$_2$Hg$_2$SF$_6$ & Ba$_2$Ca$_3$Tl$_2$Cu$_4$O$_{12}$ & 9 & $\sim$65 \\
    \bottomrule
\end{tabular}
\caption{\textbf{Sample compute times for calculating the existence of a
tie-line between two phases.} The time required is highly dependent on the
number of components, i.e.\ unique elements in the combined chemical space,
and further depends on the number of all known compounds in the chemical
space.  Each calculation was performed on a standard desktop computer
utilizing a single core.}\label{tab:sample_times_ch}
\end{table}
    
We then represent stable compounds as nodes and tie-lines as edges, thereby
generating the ``universal phase diagram'' as a complete thermodynamic phase
stability network of all inorganic materials. We use the \textsf{\textit{Qhull}}
library~\cite{Barber96thequickhull} as implemented in the \textsf{\textit{qmpy}}
package (\href{https://pypi.org/project/qmpy}{pypi.org/project/qmpy}) for all
the convex hull calculations reported in this work.\\

\noindent
\textbf{\large S2. Degree distribution of the network of all
materials}\\[-0.75\baselineskip]

\begin{figure}[!ht]
\centering
\includegraphics[width=0.55\textwidth]{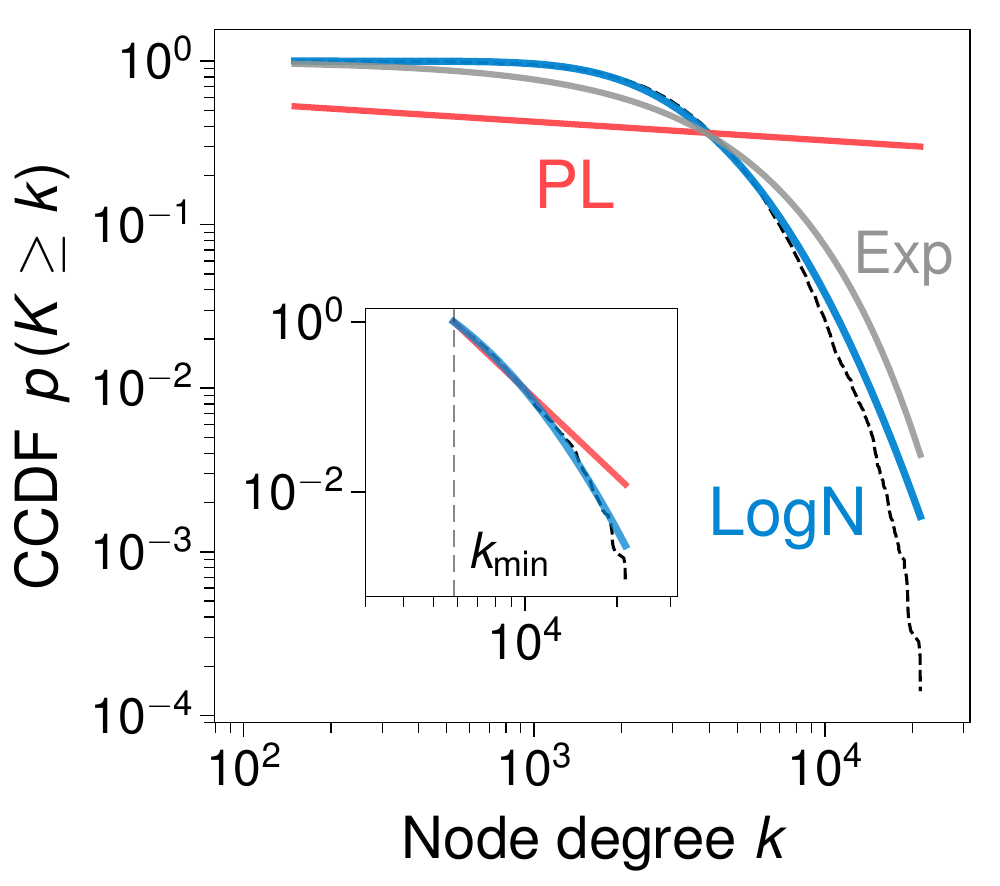}
\caption{\textbf{Fitting node connectivity data to candidate distributions.} The
complementary cumulative distribution function of the node degree in the network
of all materials is shown as dashed black lines. Power-law (PL), lognormal
(LogN), and exponential (Exp) distributions fit to the data are shown as solid
red, blue, and grey lines, respectively. The inset shows power-law and lognormal
fits to the \textit{tail} of the degree distribution for degree $k > k_{\rm min}
= 5800$.}\label{fig:s1}
\end{figure}

The probability distribution of node connectivity (number of tie-lines a
material has) in the phase stability network of all inorganic materials is
heavy-tailed. We examine which of the common heavy-tailed distributions best fit
our empirical data. In particular, several well-studied technological, social,
and biological networks are thought to have power-law distributions. Is the
thermodynamic network of materials similar to other common natural/man-made
networks exhibiting power-law behavior or not? To answer this question, we
directly compare pairs of heavy-tailed distributions using the method of log
likelihood ratios as described in Clauset~et~al.~\cite{Clauset2009}. For the
full materials network, we find that a lognormal distribution ($\mu = 8.06$,
$\sigma = 0.65$) is the best fit by far (see Fig.~\ref{fig:s1}).

We note that most empirical phenomena obey power-laws only for values greater than
some minimum value, i.e.\ only the tail of the distribution follows a power-law.
We investigate if this is indeed the case for the materials network. We find the
optimal lower-bound for a power-law behavior, $k_{\rm min}$, for the materials
network as the value that minimizes the Kolmogorov-Smirnov distance between the
data and the fit~\cite{Clauset2009}. We find $k_{\rm min}$ for the materials
network = $\sim$5800, and the power-law scaling parameter $\alpha = 4.4$. We
note that a $k_{\rm min}$ of 5800 retains only 17\% of the overall materials
network (i.e.\ only 17\% of all materials have more than 5800 tie-lines each).
Furthermore, even over this tail region of the degree distribution, a lognormal
distribution is a better fit (see the inset in Fig.~\ref{fig:s1}): the log
likelihood ratio $\mathcal{R}$ for power-law versus lognormal is $-7.15$ with a
$p$-value of 0. In other words, even in the $k_{\rm min} = 5800$ region (tail)
of the materials network, the lognormal distribution fits the data far better
than a power-law.

All analyses of fits of degree distributions mentioned above were performed with
the \textsf{\textit{powerlaw}} package~\cite{alstott_powerlaw:2014}. We note
that the graph-theoretic analyses reported in this work (e.g.\ local clustering
and centrality meaures) performed with the \textsf{\textit{graph-tool}}
package~\cite{peixoto_graph-tool_2014}, while requiring more than 8~G of memory,
take a few hours on a standard desktop utilizing up to 4 cores.

\vspace{0.5\baselineskip}
\noindent
\textbf{\large S3. New information encoded in the ``nobility
index''}\\[-0.75\baselineskip]

\begin{figure}[!ht]
\centering
\includegraphics[width=0.975\textwidth]{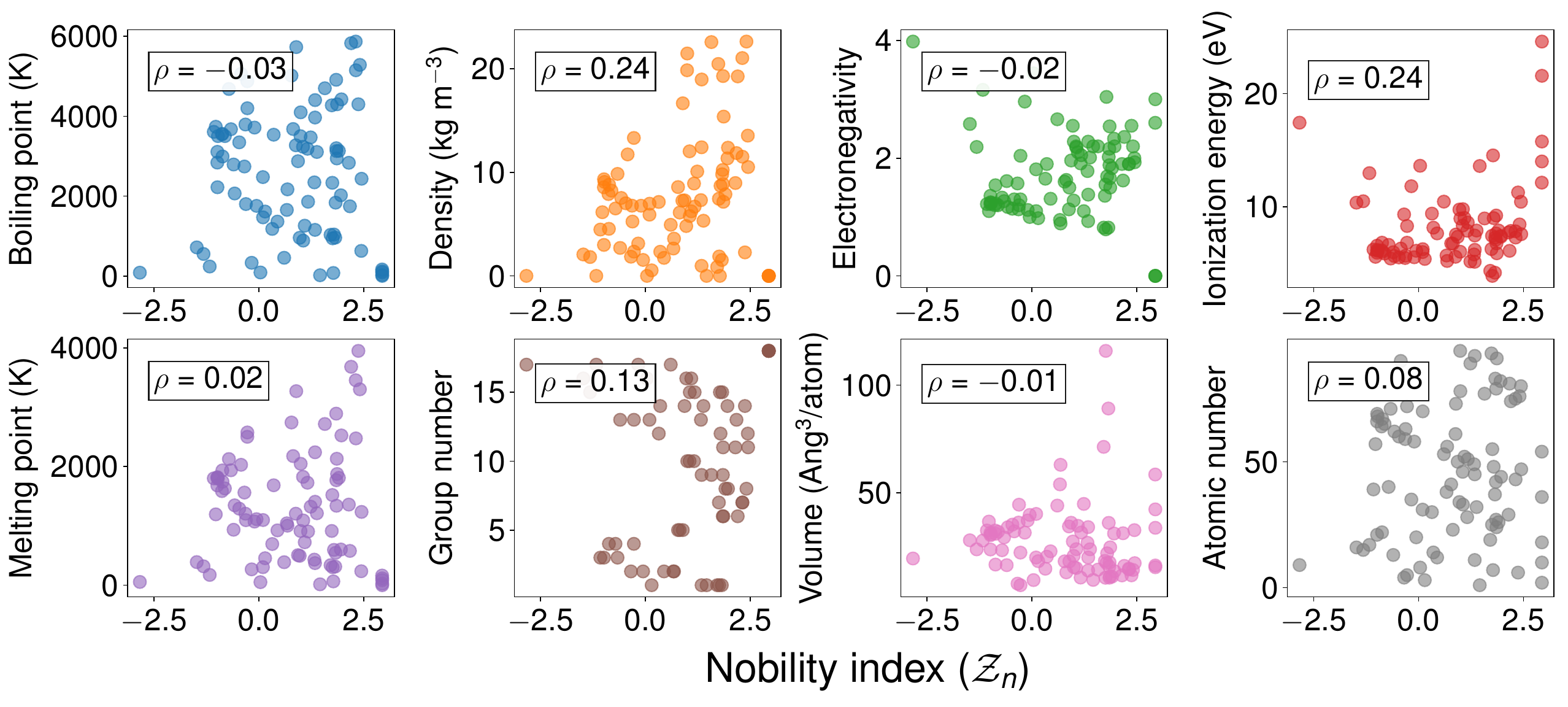}
\caption{\textbf{Comparison of nobility index versus common elemental
properties.} There is little to no correlation between the nobility index of an
element and any of its properties such as (counterclock-wise from top-left)
boiling point, density, electronegativity, first ionization energy, atomic
number, atomic volume, group in the periodic table, and melting point. The
Pearson correlation coefficient $\rho$ for each comparison is on the top-left of
the corresponding panel.}\label{fig:s2}
\end{figure}

A comparison of the nobility index $\mathcal{Z}_n$ of elements against elemental
properties such as electronegativity, boiling point, melting point, atomic
volume, etc., as collected by Ward~et~al.~\cite{Ward2015} shows little
correlation between $\mathcal{Z}_n$ and other properties, with Pearson
correlation coefficients close to 0 for most properties (see Fig.~\ref{fig:s2}
for a sample comparison set). This indicates that the nobility index defined in
this work truly encodes new information about an element/a material not
adequately captured by other common properties.

\begin{figure}[!ht]
\centering
\includegraphics[width=0.45\textwidth]{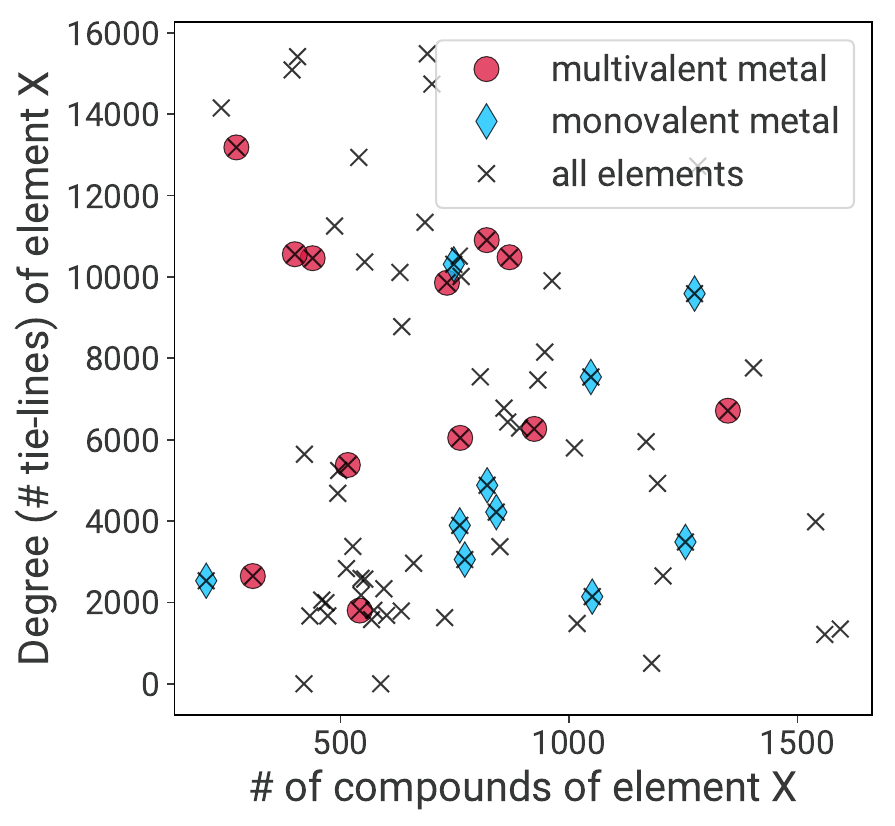}
\caption{\textbf{Comparison of number of compounds formed by an element versus its node degree.}
Multivalent metals indicated are all transition metals (Ti, V, Cr, Mn, Fe, Co,
Ni, Mo, W, Hf, Pd, Pt), and monovalent metals indicated are mostly
alkali/alkaline earth metals (Li, Na, K, Rb, Be, Mg, Ca, Sr, Al,
Zn).}\label{fig:s3}
\end{figure}

Further, data-driven metrics such as nobility index capture materials knowledge
that is not immediately intuitable or is sometimes even counter-intuitive. For
instance, intuition derived only from a few elements and some of their compounds
may imply that multivalent elements (e.g.\ transition metals) are likely to have
a higher number of tie-lines than monovalent elements (e.g. alkali metals)
simply by the virtue of a higher number of compound-forming possibilities.
However, data from all materials known so far shows no correlation between
number of compounds formed by an element and its total number of tie-lines
(i.e.\ nobility; see Fig.~\ref{fig:s3}). In fact, monovalent metals seem to
form more compounds on average than their multivalent counterparts!


\clearpage

\begin{spacing}{1.0}
\bibliographystyle{ScienceAdvances}
\bibliography{references}
\end{spacing}